\documentclass[useAMS,usenatbib,usegraphicx,letterpaper]{mn2e}

\usepackage{times}
\usepackage{color}

\title[Methanol emission in protoplanetary discs]{Towards detecting methanol emission in low-mass protoplanetary discs with ALMA: The role of non-LTE excitation.}

\author[]{Parfenov S. Yu.$^{1}$\thanks{E-mail:
Sergey.Parfenov@urfu.ru}, Semenov, D. A.$^2$, Sobolev A. M.$^1$, Gray M. D.$^3$\\
$^{1}$Ural Federal University, 51 Lenin Str., Ekaterinburg 620000, Russia\\
$^{2}$Max Planck Institute for Astronomy, K{\"{o}}nigstuhl 17, D-69117 Heidelberg, Germany\\
$^{3}$Jodrell Bank Centre for Astrophysics, School of Physics and Astronomy, University of Manchester, M13 9PL, UK\\
}

\begin{document}

\date{Accepted \today, Received \today, in original form \today}

\pagerange{\pageref{firstpage}--\pageref{lastpage}} \pubyear{2015}

\maketitle

\label{firstpage}

\begin{abstract}
The understanding of organic content of protoplanetary discs is one of the main goals of the planet formation studies. As an attempt to guide the observational searches for weak lines of complex species in discs, we modelled the (sub-)millimetre spectrum of gaseous methanol (CH$_3$OH), one of the simplest organic molecules, in the representative T~Tauri system. We used 1+1D disc physical model coupled to the gas-grain \textsc{alchemic} chemical model with and without 2D-turbulent mixing. The computed CH$_3$OH abundances along with the CH$_3$OH scheme of energy levels of ground and excited torsional states were used to produce model spectra obtained with the non-local thermodynamic equilibrium (non-LTE) 3D line radiative transfer code \textsc{lime}. We found that the modelled non-LTE intensities of the CH$_3$OH lines can be lower by factor of $>10$--$100$ than those calculated under assumption of LTE. Though population inversion occurs in the model calculations for many (sub-)millimetre transitions, it does not lead to the strong maser amplification and noticeably high line intensities. We identify the strongest CH$_3$OH (sub-)millimetre lines that could be searched for with the Atacama Large Millimeter Array (ALMA) in nearby discs. The two best candidates are the CH$_{3}$OH~$5_0-4_0~A^+$ (241.791~GHz) and $5_{-1}-4_{-1}~E$ (241.767~GHz) lines, which could possibly be detected with the $\sim5\sigma$~signal-to-noise ratio after $\sim3$~hours of integration with the full ALMA array.
\end{abstract}


\section{Introduction}
A crucial phase for planetary growth occurs in protoplanetary discs, where physical properties and chemical composition shape the emerging planetary system architecture, including primordial planetary atmospheres \citep[e.g.,][]{2011ppcd.book...55B,2013ChRv..113.9016H,2014prpl.conf..317D}. One of the most important questions related to the planet formation process is to understand how synthesis, evolution and destruction of (prebiotic) organic molecules proceed in these discs, and what fraction of these organics can reach the planetary interior and surface.

There is strong observational and laboratory evidence that the formation of complex organic molecules (COMs) begins already in cold cores of molecular clouds, prior to the onset of star formation, on dust grain surfaces serving as catalysts for mobile radicals and light atoms \citep[e.g.,][]{2009A&A...504..891O,2012A&A...541L..12B,2014ApJ...788...68O}. The newly produced molecules can eventually be returned into the gas phase either during the slow heat-up of the environment after the formation of a central star \citep[e.g.][]{2008ApJ...682..283G} or due to
other desorption mechanisms such as cosmic rays (CRP) or ultraviolet (UV) heating or directly upon surface recombination \citep[e.g.,][]{Leger_ea85,2007A&A...467.1103G}.

A direct indication that the planet-forming discs may harbour prebiotic organics comes from  a rich variety of organic compounds, including amino acids, found in carbonaceous meteorites and cometary dust in our own solar system \citep[e.g.,][]{Elsila_ea09,2012A&ARv..20...56C,2013ChRv..113.9016H,2014prpl.conf..363P}.
The first generation of organic materials could have been formed in heavily irradiated, warm regions of the presolar nebula \citep{Ehrenfreund_Charnley00,Busemann_ea06,Pizzarello_ea06,2009ARA&A..47..427H} via e.g. combustion and pyrolysis of hydrocarbons and polycyclic aromatic hydrocarbons at high temperatures, or due to X-ray/UV-induced processing of water-rich ices containing HCN or NH$_{3}$, hydrocarbons, and CO or CO$_{2}$. Then, the second generation of more complex organics could have been produced from the simpler first-generation organic matter by aqueous alteration inside large carbonaceous asteroids \citep{Ehrenfreund_Charnley00,Pizzarello_ea06}.

So far several types of organic molecules have been detected in the interstellar medium, including alcohols (e.g. CH$_3$OH), ethers (e.g. CH$_3$OCH$_3$) and acids (e.g. HCOOH), see \citet{Snyder_06,2009ARA&A..47..427H}. Recent observational results can be found in \citet{Watanabe2015} for young protostar vicinity, \citet{Crockett2015} for hot molecular core, \citet{Kalinina2010} for HII region vicinity, \citet{Kaifu2004} for dark molecular cloud. A few simple organic species such as
formaldehyde (H$_2$CO), cyclopropenylidene (c-C$_{3}$H$_{2}$), cyanoacetylene (HC$_{3}$N) and methyl cyanide (CH$_{3}$CN) have been detected and spatially resolved with (sub-)millimetre interferometers in a few nearby protoplanetary discs \citep[][]{Aikawa_ea03,Chapillon_ea12b,2013ApJ...765...34Q,2013ApJ...765L..14Q,2015Natur.520..198O}. The \textit{Spitzer} observatory has detected the infrared lines of organic molecules such as HCN and C$_2$H$_2$ in the inner, warm regions of protoplanetary discs \citep[e.g.,][]{Carr_Najita08,Salyk_ea08,2009ApJ...696..143P,2010ApJ...722L.173P,2011ApJ...733..102C,Salyk_ea11a, 2012ApJ...747...92M}.

The main reason why complex polyatomic molecules still remain largely undetected in discs is a combination of their relatively low gas-phase abundances, energy partitioning among a multitude of levels, and sensitivity limitations of observational facilities. However, deeper searches for rotational and ro-vibrational lines of complex organics in discs become possible with the Atacama Large Millimeter Array (ALMA). Therefore, it is pivotal to guide the future observational searches for weak lines of complex organics in protoplanetary discs by accurate modelling of organic chemistry and line radiative transfer in typical T~Tauri and Herbig~Ae systems.

One of the latest attempts to detect emission lines of the CH$_3$OH molecule in a young protoplanetary disc was made by \citet{vanderMarel_ea14a}. They observed the transitional disc of Oph~IRS~48 with the ALMA in Band~9 ($\sim680$~GHz) in the extended configuration during the Early Science Cycle~0. They detected and spatially resolved the H$_{2}$CO 9(1,8)--8(1,7) line at 674~GHz, which appears as a ring-like emission structure at $\sim60$~au radius. The relative H$_2$CO abundance derived with a physical disc model with a non-local thermodynamic equilibrium (non-LTE) excitation calculation is $\sim10^{-8}$. They could not detect CH$_{3}$OH emission and inferred an abundance ratio H$_2$CO/CH$_{3}$OH$\,>0.3$. \citet{vanderMarel_ea14a} predicted the line fluxes of $A$ species CH$_3$OH using their model of the Oph~IRS~48 disc, assuming non-LTE excitation. The strongest CH$_3$OH lines with the best potential for detection with the full ALMA lie within the ALMA Band~7 and Band~9.

\citet{Walsh_ea14} computed a representative T~Tauri protoplanetary disc model with a large gas-grain chemical network and found that COMs could be efficiently formed in the disc midplane via grain-surface reactions, reaching solid-state relative abundances of $\sim10^{-6}$--$10^{-4}$ with respect to H nuclei. The gas-phase COM abundances are maintained via non-thermal desorption and reach values of $\sim10^{-12}$--$10^{-7}$. Their simplified LTE line radiative transfer modelling suggests that some CH$_{3}$OH emission lines should be readily observable in nearby protoplanetary discs with the full ALMA. These lines are different from the candidate lines selected by \citet{vanderMarel_ea14a}, and have not been targeted by observations. They have an ongoing ALMA observational campaign to detect the methanol emission in large bright protoplanetary discs, but have not yet reported successful detection (Walsh, private communication).


In this study we use laminar and turbulent steady-state physical models of a typical T~Tauri disc coupled to a large gas-grain chemical model with surface synthesis of COMs (Sect.~\ref{sec:disc_phys_model}--\ref{sec:transport_model}). Using these models, and accurate methanol energy levels from CH$_{3}$OH maser modelling (Sect.~\ref{sec:coll_data}), we perform detailed non-LTE line radiative transfer (LRT) calculations of emission from the $A$ and $E$ species of methanol (Sect.~\ref{sec:lrt_calc}). In addition, we give estimates of methanol line flux densities at millimetre and sub-millimetre spectral bands of the ALMA interferometer, and present the best candidates among those methanol lines to be searched for in nearby discs with the full ALMA (50 antennas, Sect.~\ref{sec:results}). The summary and conclusions follow.

\section{Model}
\label{sec:model}

\subsection{Physical structure of the disc}
\label{sec:disc_phys_model}

In this work we adopted the flaring, steady-state 1+1D disc physical model that was computed by \citet[][hereafter, SW11]{SW11} after \citet{1999ApJ...527..893D}. This is an $\alpha$-model that represents a T~Tauri protoplanetary disc similar to that of the well-studied DM~Tau system. The gas and dust temperatures are assumed to be equal. The \citet{ShakuraSunyaev73} parametrisation of the turbulent viscosity $\nu$ was used:
\begin{equation}
 \nu(r,z) = \alpha\,c_{\rm s}(r,z)\,H(r),
 \end{equation}
where $H(r)$ is the vertical scale height at a radius $r$, $c_{\rm s}(r,z)$ is the local sound speed, and $\alpha$ is the dimensionless parameter that usually has values of $\sim0.001$--$0.1$ \citep[][]{Andrews_Williams07,Guilloteau_ea11a,2011ApJ...735..122F}.
SW11 used the constant value of $\alpha=0.01$, neglecting the potential presence of an turbulently-inactive `dead zone' \citep{1996ApJ...457..355G}.

The central star has a spectral type M0.5 with an effective temperature $T_{\rm eff}=3720$~K, a mass of $0.65~\rm{M_\odot}$, and a radius of $1.2~\rm{R_\odot}$
\citep[][]{Mazzitelli89,Simon_ea00}. The shape of the stellar UV radiation field is represented by the scaled interstellar UV radiation field of \citet{G}, with the intensity at 100~au of $\chi_*(100)=410$ \citep[][]{Bergin_ea04}. The X-ray luminosity of the star was assumed to be $10^{30}$~erg\,s$^{-1}$ (SW11).

The disc has an inner radius of 0.03~au and an outer radius of 800~au. The mass accretion rate of $\dot{M}= 4\times10^{-9}\,\rm{M_\odot}$\,yr$^{-1}$ and a disc mass of $M_{\rm disc}=0.066\,\rm{M_\odot}$ was assumed. According to the infrared observations with \textit{Spitzer} reported by \citet{Calvet_ea05,2011ApJ...732...42A,2013A&A...553A..69G}, the 5--7~Myr-old DM~Tau disc has an inner depletion of small dust at radii $3$--$10$~au. Therefore, in the chemical simulations only a disc region beyond 10~au was considered.

In the disc at $r\ga 10$~au the uniform spherical dust grain particles with a radius of $0.1\,\mu$m and amorphous olivine stoichometry were utilised. The dust grain density is $3$~g\,cm$^{-3}$ and a dust-to-gas mass ratio is a constant, $m_d/m_g=0.01$. The surface density of sites is $N_s=1.5\times 10^{15}$~sites\,cm$^{-2}$, and the total number of sites per grain is $S=1.885\times 10^6$ \citep{Bihamea01}.

\subsection{Chemical model of the disc}
\label{sec:disc_chem_model}

The chemical structure of the disc adopted in this study was computed by SW11. Their gas-grain chemical model, based on the \textsc{alchemic} code, is fully described in \citet{Semenov_ea10} and SW11. Therefore we provide only a brief summary. The chemical network is based on the osu.2007 ratefile\footnote{See: http://www.physics.ohio-state.edu/$\sim$eric/research.html} with the updates to reaction rates as of end 2010.

The 1D plane-parallel slab approach to calculating UV photo-ionisation and dissociation rates is adopted, with the stellar $\chi_*(r)=410\,(r, {\rm au})/(100~{\rm au})^{2}$ and \citet{G} interstellar $\chi_0$ UV fields that are scaled down by the visual extinction in the vertical direction and in the direction to the central star. Photoreaction rates are updated from \citet{vDea_06} (see {\it{http://home.strw.leidenuniv.nl/$\sim$ewine/photo/}}).
The self-shielding of H$_2$ from photodissociation is calculated using Eq.~(37) from \citet{DB96}. The shielding of CO by dust grains, H$_2$, and its self-shielding is calculated using the precomputed table of \citet[][Table~11]{Lea96}.

The attenuation of CRP is calculated using Eq.~(3) from \citet{Red2}, using the standard ionisation rate of $1.3\times10^{-17}$~s$^{-1}$.
Ionisation due to the decay of short-lived radionuclides is also taken into account, with the rate of $6.5\times10^{-19}$~s$^{-1}$ \citep{FG97}. The attenuation of the stellar X-ray radiation is modelled using the approximate expressions from the 2D Monte-Carlo simulations of \citet{zetaxa,zetaxb}.

The gas-grain interactions consist of adsorption of neutral species and electrons to dust grains with $100\%$ probability, and desorption of ices by thermal, CRP-, and UV-induced desorption mechanisms. In addition, dissociative recombination and radiative neutralisation of ions on charged grains and grain re-charging are taken into account. H$_2$ is not allowed to stick to grains because the binding energy of H$_2$ to pure H$_2$ mantle is low, $\sim 100$~K \citep{Lee1972}, and it freezes out in substantial quantities only at temperatures below $\sim 4$~K. Chemisorption of surface molecules with a probability of $5\%$ is considered. The UV photodesorption yield of $10^{-3}$ is assumed \citep{Oeberg_ea07,Oeberg_ea09a,Oeberg_ea09b,Fayolle_ea11a}.

An extended list of surface reactions along with desorption energies and a list of ice photodissociation reactions is taken from \citet{Garrod_Herbst06}. The standard rate equation approach without H and H$_2$ tunnelling is considered. Overall, the disc chemical network consists of 657 species made of 13 elements, and 7\,306 reactions. The `low metals' initial abundances of \citet{Lea98} are utilised, see
Table~\ref{tab:inabun}.

\begin{table}
\caption{Initial abundances used by SW11 for computations of the disc chemical evolution models adopted in this study.}
\label{tab:inabun}
\begin{tabular}{@{}ll}
\hline
Species & Abundance relative to the total \\
        & number of H nuclei\\
\hline
H$_2$&   $0.499$     \\
H    &   $2.00(-3)$  \\
He   &   $9.75(-2)$  \\
C    &   $7.86(-5)$  \\
N    &   $2.47(-5)$  \\
O    &   $1.80(-4)$  \\
S    &   $9.14(-8)$  \\
Si   &   $9.74(-9)$  \\
Na   &   $2.25(-9)$  \\
Mg   &   $1.09(-8)$  \\
Fe   &   $2.74(-9)$  \\
P    &   $2.16(-10)$ \\
Cl   &   $1.00(-9)$  \\
\hline
Note. $a(b)=a\times10^b$
\end{tabular}
\end{table}

\subsection{Turbulent chemistry model and methanol abundances}
\label{sec:transport_model}
The gas-phase species and dust grains are assumed to be perfectly mixed and transported with the same diffusion coefficient
\begin{equation}
D_{\rm turb}(r,z) = \nu(r,z)/Sc.
\end{equation}
Here, $Sc$ is the Schmidt number describing the efficiency of turbulent transport and is treated as a free parameter. In this study we considered the two extreme cases, namely, the `laminar' disc model without turbulent transport with $D_{\rm turb}=0$ ($Sc = \infty $) and the disc model with `fast' turbulent mixing ($Sc=1$). These approximations were commonly used in other disc chemical studies, \citet[see, e.g.,][]{2004A&A...415..643I, Willacy_ea06, Heinzeller_ea11, Furuya_ea13}. No inward and outward diffusion across boundaries of the disc and no transport through the disc midplane were allowed.

\begin{figure*}
\centering
\includegraphics[scale=0.7]{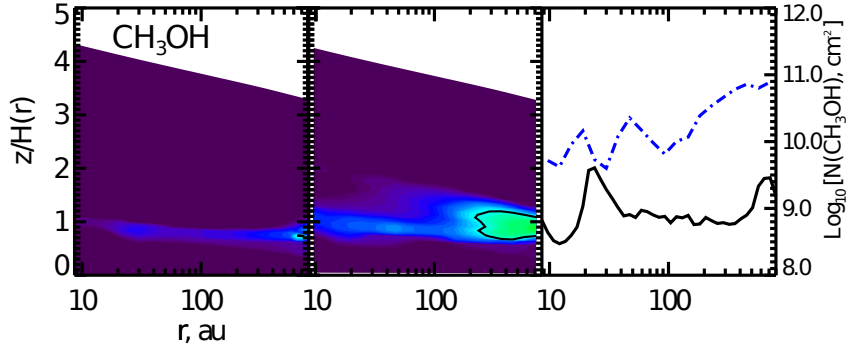}
\caption{The $\log$ of relative abundances with respect to the total hydrogen
density (first 2 panels) and vertical column densities (right panel) of gaseous methanol in the DM~Tau disc at 5~Myr are presented. (From left to right) Results for the two turbulence transport models are shown: (1) the `laminar' DM~Tau disc model and (2) the `fast' 2D-mixing DM~Tau disc model. The contour line denotes the relative CH$_{3}$OH abundance of $10^{-12}$ relative to H nuclei.
The solid line in the right panel corresponds to the `laminar' disc case, while the dash-dotted represents the `fast' 2D-mixing disc case. The figure is taken from SW11.}
\label{fig:methanol_abunds}
\end{figure*}

Using the disc physical and turbulent chemistry models, and initial abundances described above, SW11 calculated the time-dependent abundances over the time span of 5~Myr. In Fig.~\ref{fig:methanol_abunds} the calculated abundances and column densities of the gas-phase methanol at 5~Myr in the DM~Tau disc are shown. These abundances were used in our study for LRT calculations.

As can be clearly seen in Fig.~\ref{fig:methanol_abunds}, methanol in both disc models is distributed predominantly above the disc midplane, where the kinetic temperatures are about $30$--$50$~K. The turbulent transport enhances the gas-phase methanol abundances and column densities by more than one order of magnitude, particularly in outer disc region. This is due to the more efficient formation of methanol on dust grain surfaces in the transport model. There icy grains from the cold dark midplane can reach warmer or more heavily irradiated disc regions, which facilitate more intense photoprocessing of simple water-rich ice mantles and formation of reactive radicals, and increased surface mobility of these heavy species (SW11). In turn, higher abundances of solid methanol naturally lead to enhanced concentrations of gas-phase methanol
via thermal and non-thermal (photo) desorption. Please also bear in mind that the methanol molecular layer appears at various vertical heights in the two models, which can have strong impact on the results of the line radiative transfer modelling (see below).

In the chemical simulations by SW11, the methanol has been considered as a single species, without distinguishing between the $A$ and $E$ species that represent various configurations of the nuclear spin states of the methyl group. For our LRT calculations we assumed that the $A$ and $E$ species of methanol are equally abundant \citep[see e.g.][]{Flower2010}.

Intrinsic uncertainties of molecular column densities, including CH$_3$OH,  predicted by modern disc chemical models can be as large as factor of $3$--$5$ due to the intrinsic uncertainties in the reaction rates in the chemical networks \citep[e.g.,][]{Vasyunin_ea08}. To take these uncertainties into account when computing the intensities of the strongest methanol lines, we considered an additional case where the `laminar' methanol abundances were multiplied by an additional factor of 5.

\subsection{Disc structure used in LRT calculations}
\label{sec:disc_model_lrt}
SW11 performed their calculations with a two-dimensional non-regular grid consisting of 41 radial and 91 vertical points in cylindrical coordinates ($r$, $z$). The distance between the grid points along the vertical axis $z$ is smaller for lower $r$. For example, at $r=9.66$~au the grid spans the vertical range from 0.054 to 4.82~au, while at $r=800$~au the vertical range is from 17.3 to $1\,536$~au.

For line radiative transfer calculations we had to use another adaptive 3D grid (see Sect.~\ref{sec:lrt_calc}), and applied linear interpolation to the disc parameters from the physical and chemical grid of SW11. The disc physical parameters are the molecular hydrogen number density $n_{\rm{H2}}$, the helium number density $n_{\rm{He}}$, gas kinetic temperature $T$, and the methanol abundance $X_{\rm{M}}$ (the abundance of $A$ or $E$ species methanol relative to the H$_2$ number density). For the LRT computations, the disc physical parameters across the disc midplane at $z=0$ were taken to be equal to the parameters at the lowest vertical grid cell. The ranges of the disc physical parameters are given in Table~\ref{tab:par_minmax}.

\begin{table}
\caption{The minimum and maximum values of physical parameters in the SW11 disc models adopted in this study.}
\label{tab:par_minmax}
\begin{tabular}{@{}lccc}
\hline
Disc model & Parameter & Minimum & Maximum\\ \hline
`Laminar' and `fast' & $T$, K & 12 & 114\\
mixing models & &  & \\
`Laminar' and `fast' & $n_{\rm{H2}}$, cm$^{-3}$ & $1.63(-7)$ & $1.26(11)$\\
mixing models & &  & \\
`Laminar' and `fast' & $n_{\rm{He}}$, cm$^{-3}$ & $6.66(0)$ & $2.45(10)$\\
mixing models & &  & \\
`Laminar' model & $X_{\rm{M}}$ & $5.98(-31)$ & $1.48(-12)$\\
`Fast' mixing model & $X_{\rm{M}}$ & $1.61(-32)$ & $4.08(-11)$\\
\hline
Note. $a(b)=a\times10^b$
\end{tabular}
\end{table}

\subsection{Methanol energy levels scheme, radiative and collisional rate data}
\label{sec:coll_data}
In this study we used the scheme of methanol levels representing a subset of the larger set of rotationally and torsionally excited levels of the methanol ground vibrational state described in \citet{Cragg_ea05}. The scheme of \citet{Cragg_ea05} was used for calculations of the level population numbers in the medium with gas kinetic and dust temperatures in the range from 20 to 250~K, hydrogen number densities from $10^4$ to $10^9$~cm$^{-3}$, and specific column densities from $10^{10}$ to $10^{14}$~cm$^{-3}$~s. So, the scheme was approved for calculations in the range of physical parameters corresponding to conditions in the disc models considered in this paper.
The method used in our study to minimize the number of levels while retaining accuracy of the LRT modelling was proposed by \citet{Sobolev94}. Briefly, the rotational levels of the methanol ground torsional state are included up to a pre-selected energy threshold. The levels of the torsionally excited states are included when they are coupled with the selected ground levels by radiative transitions. In our line transfer calculations we considered the ground state levels with energies up to 285~cm$^{-1}$ (or 400~K) and with rotational angular momentum quantum number $J$ up to 18. Hereafter, we will designate the scheme including only levels of the ground torsional state as the `$v_t=0$' scheme. This scheme includes 190 and 187 levels for $A$ and $E$ species CH$_3$OH, respectively.

Most of our results are obtained using the `$v_t=0$' scheme as it includes a relatively low number of levels. This scheme has moderate computational demands in terms of memory and CPU time. We also performed computations with the other level scheme which includes
rotational levels of the first two torsionally excited states with torsional quantum numbers $v_t=1$ and $v_t=2$. Hereafter, we will designate this larger set of levels as the `$v_t=2$' scheme. The total number of levels in the `$v_t=2$' scheme is 570 and 561 for $A$ and $E$ species methanol, respectively. The maximum energy of the levels in the `$v_t=2$' scheme is about 850~cm$^{-1}$ or $1\,200$~K.

The level energies and radiative transition rates used in our study were computed by \citet{Cragg_ea05} using the data of \citet{Mekhtiev99}. The collisional rates were taken from the data of \citet{Rabli2010MNRAS40695R,Rabli2011MNRAS4112093R} that are available online\footnote{http://massey.dur.ac.uk/drf/methanol\_H2/, http://massey.dur.ac.uk/drf/methanol\_He/} and are the same as used in the Leiden Atomic and Molecular Database (LAMDA, \citet{lamda}). For our calculations these collisional rates were scaled according to the level energies in the same manner as was done by \citet{Cragg_ea05}. The rates for the pure rotational transitions missing in the \citet{Rabli2010MNRAS40695R,Rabli2011MNRAS4112093R} data were obtained using the propensity rules described in \citet{Cragg_ea05}. The rates for the transitions between torsional states were assumed constant and equal to $5\times10^{-14}$~cm$^3$~s$^{-1}$ \citep{Cragg_ea05}. In the computations with our `$v_t=0$' scheme, we considered collisions with para-H$_2$ only, as implemented in the data file from the LAMDA database. In the computations with the `$v_t=2$' scheme we included both para-H$_2$ and helium as collisional partners. We did not consider collisions with ortho-H$_2$ as the corresponding data were not openly available at the moment when most of our LRT computations were done.

In summary, the main difference between the LAMDA methanol molecular rate data and our `$v_t=0$' data are the values of the Einstein A-coefficients and scaling factors applied to the collisional rates according to the energy of the levels. To verify how much this difference in the methanol molecular data affects the results of our LRT modelling, we computed the CH$_3$OH spectra using the molecular data from the LAMDA database and compared it with the spectra obtained with the `$v_t=0$' scheme.

\subsection{Non-LTE radiative transfer setup}
\label{sec:lrt_calc}
The line radiative transfer calculations with the adopted disc physical and chemical models were performed with the LIne Modeling Engine (\textsc{lime}), version 1.41b \citep{2010A&A...523A..25B}. \textsc{lime} is a 3D non-LTE Monte-Carlo radiative transfer code designed for modelling (sub-)millimetre and far-infrared continuum and spectral line radiation. \textsc{lime} performs radiative transfer calculations using a weighted sample of randomly selected grid points in 3D space that is represented by a spherical computational domain.

To create the weighted sample of 3D grid points, using 1+1/2D disc grid described in Section~\ref{sec:disc_model_lrt}, we assumed a mirror symmetry with respect to the disc midplane and an azimuthal symmetry of the disc. The weighted sample of grid points was selected according to the following rules similar to those used by \citet{Douglas_ea13}.

First, a grid cell is randomly selected within the spherical computational domain that has a radius of 1\,600~au (twice the disc radius). Then, this grid point becomes a candidate to be included in the \textsc{lime} grid. If this grid point is not located within the disc boundaries, then it is rejected and the whole process is restarted. If the point is located within the disc boundaries, then a random number $p$ is generated in the interval $[0,1]$ and the gas density $n_{\rm{H2}}$ and methanol abundance $X_{\rm{M}}$ are computed at that grid point. The point is included in the sample only when $p<\left( n_{\rm{H2}}/n_{\rm{0H2}} \right)^{0.3}$ and $p<\left( X_{\rm{M}}/X_{\rm{0M}} \right)^{0.3}$, where $n_{\rm{0H2}}$ and $X_{\rm{0M}}$ are the maximum values of the H$_2$ density and methanol abundance in the disc, respectively (see Table~\ref{tab:par_minmax}). After the weighted sample of points is selected it is smoothed. The smoothing method was the same as described by \citet{2010A&A...523A..25B}, but we modified it so that the all grid points remain within the boundaries of our disc model.

\begin{figure*}
\centering
\includegraphics[scale=0.65]{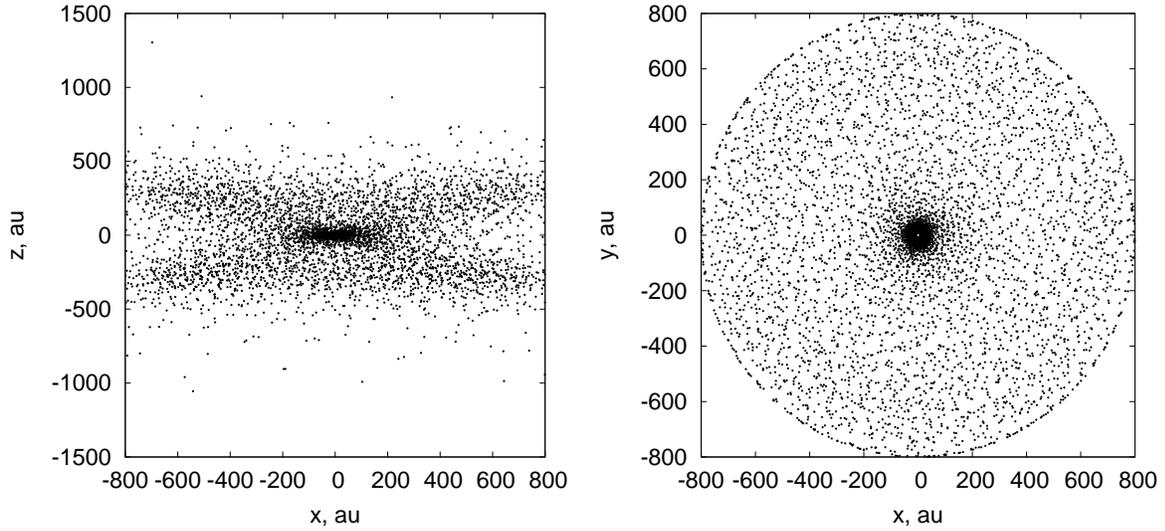}
\caption{The `laminar' disc grid used to perform radiative transfer calculations with the
\textsc{lime} code. (Left panel) the disc model has an edge-on orientation. (Right panel)
the disc is seen face-on.}
\label{fig:small_grid}
\end{figure*}

To trace the photons that leave the computational domain, \textsc{lime} uses another grid of sink points that are uniformly distributed over the surface of the computational domain. In our LRT simulations we used 10\,000 grid points for the disc model and 10\,000 sink points. For example, the grid generated for the `laminar' disc model is presented in Fig.~\ref{fig:small_grid}. All results presented below for the `laminar' model were obtained with this grid. This grid is denser in central disc regions, with $r<100$~au, where $n_{\rm{H2}}$ is high. In accordance with CH$_3$OH spatial distribution, the number of grid points in the disc midplane at $r>200$~au is lower than in regions above the midplane with $z>100$~au.

The \textsc{lime} LRT modelling results depend on the selected parameters that control the computational grid, e.g., the total number of grid cells and number of sink points. Due to intrinsic randomness when constructing a \textsc{lime} grid, there can be a difference between synthetic spectra computed with different grids even when those are generated with the same parameters. We treat such difference as an additional uncertainty in the modelled spectra. To quantify this uncertainty, we compared the methanol spectra computed with different \textsc{lime} grids, but generated with the same grid point selection rules and disc parameters (see Appendix~A). In addition, to verify the choice of our values for the total number of 3D \textsc{lime} grid points, and the number of \textsc{lime} sink points, on the results of our LRT modelling, we also performed calculations with two denser grids (see Appendix~A). The first grid includes 20\,000 points for the disc model and 20\,000 sink points. The second grid includes 40000 points for the disc model and 40\,000 sink points. The sample of points in the second grid was created from the first 20000-point grid by adding 20\,000 additional points. The positions of the additional 20\,000 points were generated with the same rules as described above but the random number $p$ was compared only with $\left( X_{\rm{M}}/X_{\rm{0M}} \right)^{0.3}$ value. Thus, the 40\,000-point grid is relatively dense in the disc regions with high methanol abundance.

As a kinematic model, we assumed Keplerian rotation around a star with the mass of 0.65~$\rm{M}_{\sun}$, and a uniform micro-turbulent velocity component with the line-width of 0.1~km~s$^{-1}$. This turbulent velocity is close to the values inferred from observations of protoplanetary discs \citep[see, e.g.,][]{2012A&A...548A..70G,deGregorio_Monsalvo_ea13,2014prpl.conf..317D}.

The dust opacity was taken from \citet{OH94}. In this study we considered the opacity computed for dust grains with thin initial ice mantles and a gas number density of $10^6$ cm$^{-3}$. We also made test calculations with the dust opacity modelled for grains with thin initial ice mantles and a larger gas number density of $10^8$ cm$^{-3}$, and found that our results were not altered significantly by changes in the dust opacity law.

Computations were performed independently for the $A$ and $E$ species of methanol, as they were not considered to be coupled. To compute the level populations with the \textsc{lime} code we used 20 global iterations and more than $10^6$ photons. This results in a relatively low residual Monte-Carlo noise level in the synthetic spectra and level populations (see Appendix~A for more details).

\section{Results}
\label{sec:results}
The outcomes of our LRT modelling are synthetic disc images and spectra in various methanol lines. The spectra are computed as the sum of emission intensities in all the pixels of the disc image. The line flux densities are obtained by subtracting dust thermal continuum emission from the disc spectra. The standard spectra presented in this study are obtained assuming that the disc is located at the distance of 140~pc and a disc inclination angle of $35\degr$ ($0\degr$ is the face-on orientation). The distance and orientation of the disc are similar to those for a DM~Tau disc \citep{2011ApJ...732...42A}. The size of a pixel in the synthetic images is 0.01~arcsec. This size is close to the beam size achievable in the most extended ALMA configuration (6~mas) at 675~GHz. Note that the difference between spectra computed with pixel sizes that differ by factor of several from our 0.01~arcsec value does not exceed the Monte-Carlo noise level (see Appendix~A for the Monte-Carlo noise levels). All spectra presented in this study were obtained with a velocity resolution of 0.2~km~s$^{-1}$.

\begin{figure*}
\centering
\includegraphics[scale=0.56,angle=270]{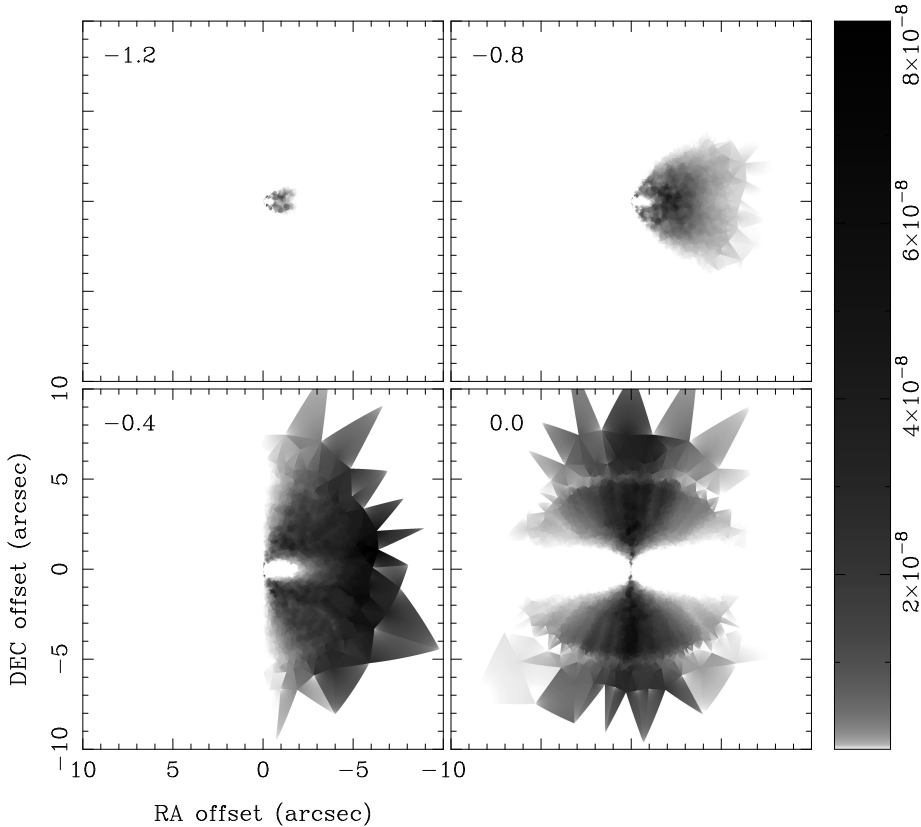}
\caption{The channel map of the $6_0-5_0~A^+$ (290.111~GHz) CH$_3$OH emission line. The methanol
abundances are from the `fast' mixing disc chemical model.
The intensity is expressed in units of Jy/pixel. The label at the left top corner of each
panel is a velocity shift in km~s$^{-1}$ from 290.111~GHz.}
\label{fig:disc}
\end{figure*}

An example of the calculated channel map of the $A$ species CH$_3$OH emission line at 290.111~GHz is presented in Fig.~\ref{fig:disc}. We found that this is one of the strongest lines in the synthetic spectra computed with the `$v_t=0$' methanol level scheme, using the `fast' mixing disc chemical model. In this image one can see artefacts due to the relatively low number of \textsc{lime} grid points sampling outer parts of the disc model. These points make the largest contribution to the uncertainty in the synthetic spectra. For the parameters of \textsc{lime} grids used in our study, this uncertainty is relatively large and can reach about 30$\%$ of the line absolute flux. However, this is still comparable or lower than possible bias due to uncertainties in chemical reaction rates and molecular transitions rates.

\subsection{`Laminar' disc chemical model}
\label{sec:res:laminar}
In this Section we present the \textsc{lime} LRT modelling results obtained with the `$v_t=0$' methanol level scheme and the `laminar' disc chemical model. The `laminar' disc model has very low methanol abundances $X_{\rm{M}}\la 10^{-12}$ (wrt the total amount of hydrogen nuclei, see Fig.~\ref{fig:methanol_abunds}) and serves as the pessimistic case for predicting the intensities of methanol lines searchable in the ALMA bands.

The Monte-Carlo noise level of the non-LTE spectra computed for the `laminar' disc model is about 0.01~mJy. The uncertainty in absolute line flux densities related to the realisation of \textsc{lime} grid-point distribution is of order 0.2~mJy (see Appendix~A). The strongest lines in the non-LTE LRT calculations are from the $J_1\rightarrow{}J_0$ and $J_0\rightarrow{}J_0$ $A$ species CH$_3$OH line series, and from the $J_{-1}\rightarrow{}J_{-1}$ $E$ species CH$_3$OH line series.

The peak line flux densities computed with `$v_t=0$' methanol level scheme are presented in Fig.~\ref{fig:nlte_vs_lte_laminar}. In Fig.~\ref{fig:nlte_vs_lte_laminar} one can see that both the LTE and non-LTE line fluxes computed with the `laminar' disc model are rather low and do not exceed 4~mJy. The peak flux densities are higher by a factor of 3.5 than those shown in Fig.~\ref{fig:nlte_vs_lte_laminar} if the disc has a perfect face-on orientation. Anyway, for any disc inclination the CH$_3$OH lines are undetectable even within 9 hours of on-source integration with the `Full Science' ALMA capabilities, taking into account the 0.2~mJy uncertainty in the absolute values of the synthetic line flux densities.

We also computed the LTE and non-LTE spectra using the `laminar' disc model with methanol abundances that are scaled up by a factor of 5. The intensities of the LTE and non-LTE spectra are higher by the same factor of 5 compared to the spectra presented in Fig.~\ref{fig:nlte_vs_lte_laminar}. This means that the line flux density scales linearly with methanol abundance, as the line optical depth is very low and does not exceed 0.05 for any disc inclination. Consequently, abundance uncertainties predicted by the chemical simulations translate directly into the uncertainty in the \textsc{lime} flux densities.

Although the line flux densities predicted with the `laminar' disc model may be too pessimistically low, we use this disc model to demonstrate the difference between the non-LTE and LTE methanol spectra. We also use this `laminar' model to demonstrate the changes in the computed methanol spectra due to variation of the radiative and collisional rate-coefficient data in the molecular model.

\subsubsection{The importance of non-LTE effects}
\begin{figure*}
\centering
\includegraphics[scale=0.8]{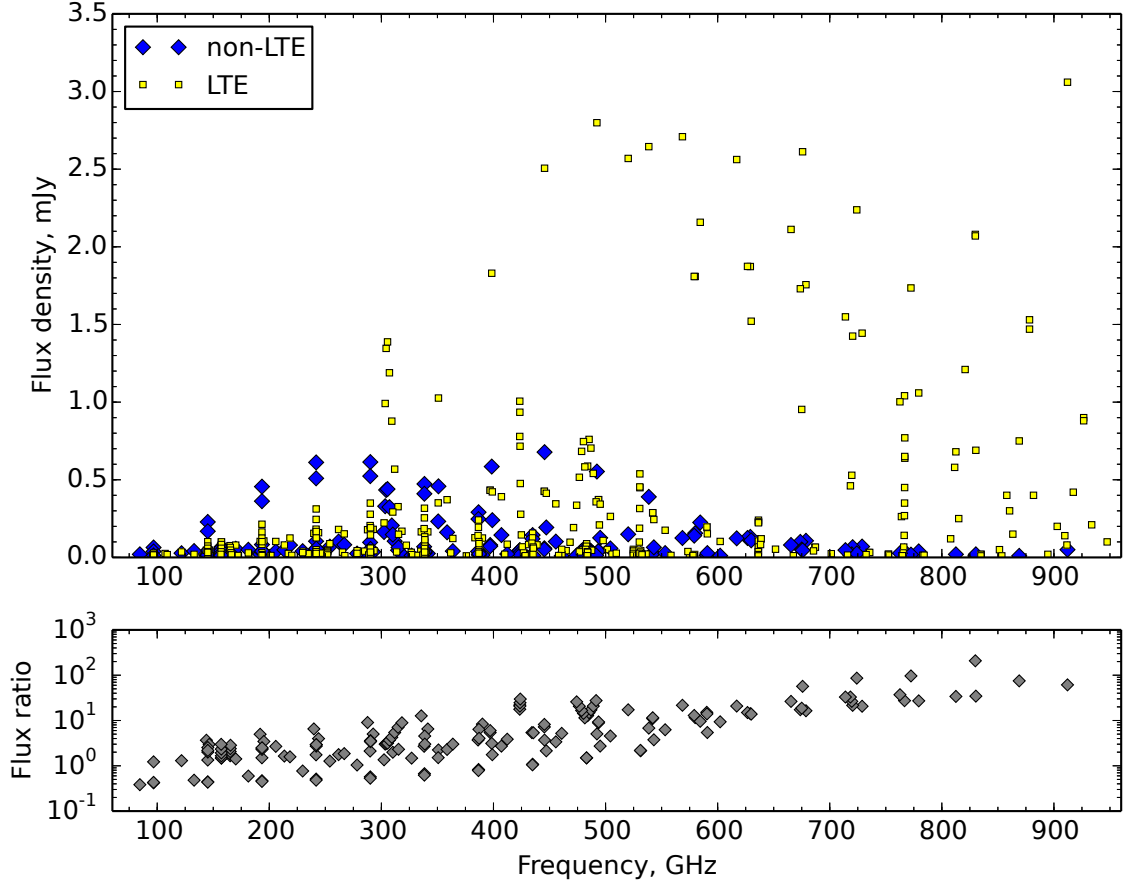}
\caption{The line peak flux densities of methanol obtained under LTE and non-LTE assumptions for the disc located at the distance of 140~pc and inclined by $35\degr$.
The methanol abundances are from the `laminar' disc model.
The `$v_t=0$' methanol level scheme is used.
Only lines with the peak flux densities $>0.01$~mJy are shown.}\label{fig:nlte_vs_lte_laminar}
\end{figure*}

To show the importance of non-LTE effects for modelling methanol lines, we compare the spectra computed with `$v_t=0$' scheme assuming LTE and non-LTE excitation. As can be clearly seen in Fig.~\ref{fig:nlte_vs_lte_laminar}, the non-LTE line peak flux densities are in general lower than their LTE counterparts. Many lines predicted to be strong by the LTE calculations appear relatively weak in the non-LTE case. There is a  number of lines lying at frequencies $>600$~GHz for which the difference between LTE and non-LTE flux densities exceeds the factor of 10. For example, non-LTE flux density predicted for the line at 675.773 GHz ($3_3-2_2~E$ transition) is lower by a factor of 60 than the LTE flux density. Such differences between non-LTE and LTE flux densities are comparable with the total intrinsic uncertainties of the calculated methanol abundances in SW11 models adopted in our study (see Sections \ref{sec:transport_model} and \ref{sec:diss:chem_model}). The strongest methanol lines in the non-LTE spectrum lie in the frequency range from 200 to 600~GHz, which corresponds to the ALMA Bands from 6 to 8. Our LTE spectrum is well consistent with the methanol spectrum computed by \citet{Walsh_ea14} (in terms of relative line flux densities).

\begin{figure}
\centering
\includegraphics[scale=0.46]{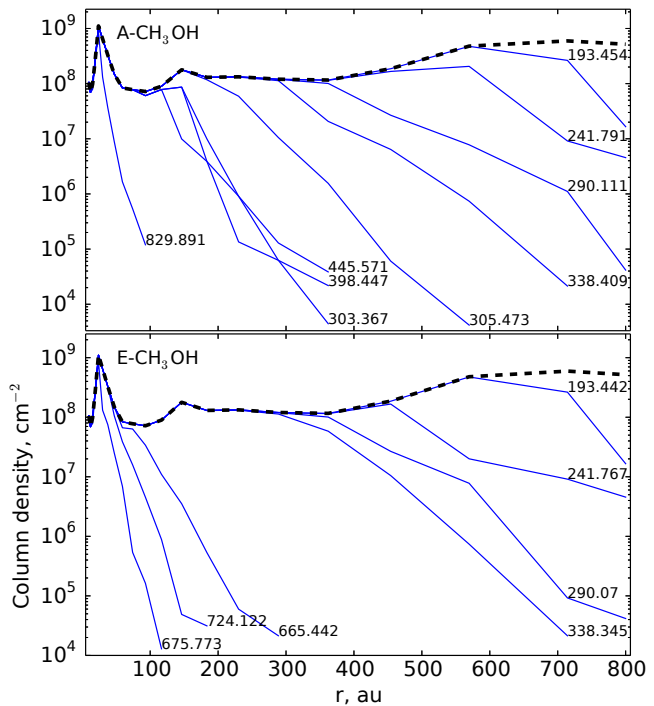}
\caption{The critical column density diagrams for $A$ species (upper panel) and $E$ species (lower panel) methanol transitions obtained for the `laminar' disc model. The radial distribution of the total methanol column density is shown with dashed line. The column density in the disc regions with supercritical density for a given transition is shown as a solid line labeled with the transition frequency in GHz.}\label{fig:CCD}
\end{figure}

The LTE holds in the case of high transition optical depth or when molecular levels are populated mostly by collisions and thus are thermalized. The latter occurs when the hydrogen density is significantly higher than the critical density of a transition $n_{\rm{cr}}$. The line of sight optical depths are small in our models (see Section~\ref{sec:res:laminar}). Thus, the main reason for deviations from LTE in our spectra is that levels are not thermalized. This can be demonstrated with critical column density diagrams, i.e. the radial distribution of hydrogen column density in the sub- and supercritical disc regions \citep{Pavlyuchenkov2007}. The sub- and supercritical disc regions are regions with the hydrogen density that is, respectively, lower and higher than the thermalization density $n_{\rm{th}}=10 n_{\rm{cr}}$ at which level populations become populated mainly by collisions \citep{Pavlyuchenkov2007}. It is seen in Fig.~\ref{fig:CCD} that the strongest transitions are not completely thermalized. The supercritical disc region tends to shrink with increasing transition frequency. This can explain why the difference between LTE and non-LTE flux densities increases with increasing transition frequency (see lower panel in Fig.~\ref{fig:nlte_vs_lte_laminar}).

Population inversion occurs in some methanol transitions in our disc models. Among these transitions the strongest belong to $J_2\rightarrow{}J_1~E$ (the strongest transition is $5_2-4_1~E$), $J_{-1}\rightarrow{}J_0~E$ (the strongest transition is $7_{-1}-6_0~E$), $J_{3}\rightarrow{}J_2~A^+$ and $A^-$ methanol series (the strongest transitions are $3_{3}-3_2~A^+$ and $A^-$). However, these inversions do not lead to considerable maser amplification and formation of intense maser emission. The line of sight optical depths remain positive in all considered transitions.

\subsubsection{The importance of different methanol molecular data}

The deviations from LTE are strong in our model. Thus the molecular level scheme and data used for LRT calculations become important for feasible calculations of line fluxes. To estimate the influence of excited methanol levels and collisions with helium we computed and compared the spectra using our `$v_t=0$' and `$v_t=2$' methanol level schemes.

\begin{figure}
\centering
\includegraphics[scale=0.46]{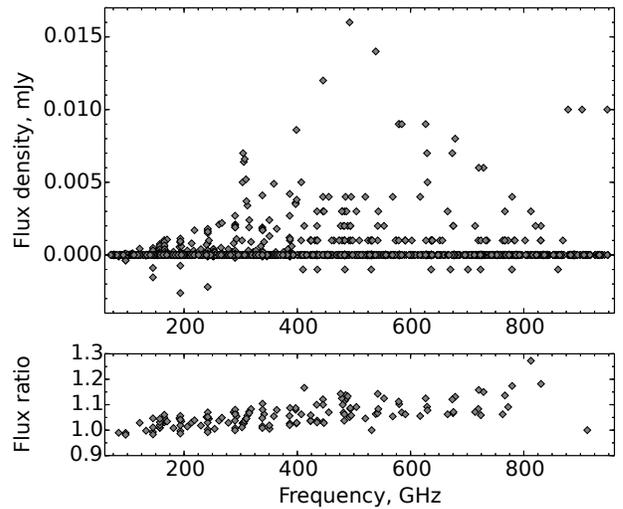}
\caption{The difference (upper panel) and ratio (lower panel) between the line peak flux
densities obtained with `$v_t=2$' and with `$v_t=0$' CH$_3$OH level schemes. The difference
is shown only for the transitions that are present in the `$v_t=0$' scheme. The ratio is shown
only for the lines with peak flux densities $>0.01$~mJy.}
\label{fig:vt0_vt2}
\end{figure}

The difference between these two spectra is presented in Fig.~\ref{fig:vt0_vt2}. Only transitions that are present in the `$v_t=0$' scheme are shown. The difference is slightly higher than the Monte-Carlo noise level of 0.01~mJy, and the flux densities of the remaining lines in the `$v_t=2$' scheme do not exceed 0.02~mJy. From the bottom panel in Fig.~\ref{fig:vt0_vt2} one can see that the high-frequency lines ($\ga 300$~GHz) are systematically more intense compared to the low-frequency lines when the `$v_t=2$' scheme is used. This effect cannot be due to Monte-Carlo noise and is not due to the assumed \textsc{lime} grid parameters as the grid was identical in both models. This difference is relatively small and arises due to the inclusion of collisions with helium atoms.

Thus, the methanol level scheme and collisional data similar to those from the LAMDA database, allow us to obtain reliable estimates of the methanol line flux densities, at least for the adopted disc model. The inclusion of collisions with He significantly affects the line ratios, particularly for those transitions with energy levels separated by more than $\sim 200$~GHz.

\begin{figure}
\centering
\includegraphics[scale=0.46]{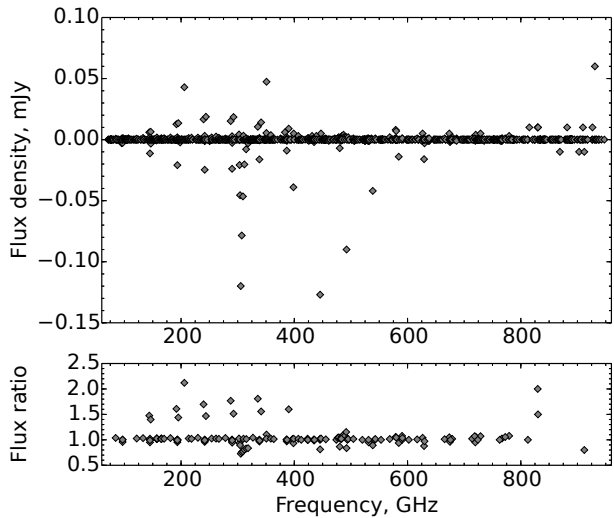}
\caption{The difference (upper panel) and ratio (lower panel) between the line peak flux
densities obtained with the LAMDA and with `$v_t=0$' methanol level schemes. The difference is shown only for the transitions that are present in the `$v_t=0$' scheme. The ratio is shown only for the lines with peak flux densities $>0.01$~mJy. The $x$-axis is the line frequencies taken from the LAMDA database. Note that the transition frequencies are different in our and LAMDA data (the mean and maximum difference is $\sim 80$ and 350 kHz, respectively).}
\label{fig:vt0_vs_lamda}
\end{figure}

In Fig.~\ref{fig:vt0_vs_lamda} we present the difference and ratio between the methanol spectrum computed with the level scheme and molecular data taken from the LAMDA database and the spectrum computed with our `$v_t=0$' CH$_3$OH level scheme. It is clear that the difference between line flux densities obtained with the LAMDA and our `$v_t=0$' methanol level schemes does not exceed the Monte-Carlo noise level of 0.01~mJy for the most of transitions.

\begin{figure}
\centering
\includegraphics[scale=0.46]{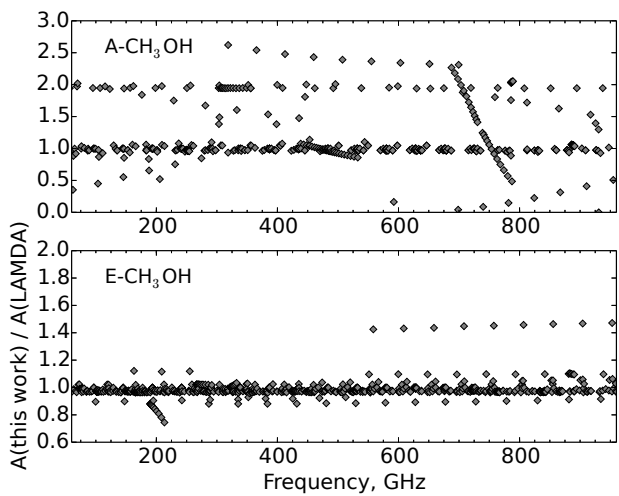}
\caption{The ratio of Einstein A-coefficients used in this work and A-coefficients taken from the LAMDA database for the $A$ and $E$ species methanol. The ratio is shown only for the transitions which are present in the `$v_t=0$' scheme.}
\label{fig:A_ratio}
\end{figure}

However, for a number of transitions alteration of molecular data results in the flux changes $>10\%$. In these cases, the difference comes from the difference in the Einstein A-coefficients adopted in the LAMDA database and our `$v_t=0$' level scheme. The A-coefficients in the LAMDA database were computed using the data from the JPL database, while we used the data obtained by \citet{Mekhtiev99}. The A-coefficients used by us are almost identical to those from CDMS database \citep{Muller2001,Muller2005}. For some transitions our A-coefficients and A-coefficients from the LAMDA database differ by more than a factor of 2 (see Fig.~\ref{fig:A_ratio}). It is hard to find the source of these deviations, which could simply be data errors.

\subsection{`Fast' mixing disc chemical model}
\label{sec:fastmix}

In this Section we present the \textsc{lime} LRT modelling results obtained with the `$v_t=0$' methanol level scheme and the `fast' mixing disc chemical model. The `fast' mixing disc model has gas-phase methanol abundances higher by about an order of magnitude than the `laminar' model (see Fig.~\ref{fig:methanol_abunds}). We consider it as the optimistic case for simulations of the methanol line intensities searchable within the ALMA bands.

The spectrum computed for the `fast' mixing model is presented in Fig.~\ref{fig:fastmix}. The Monte-Carlo noise level of the spectra is about 0.02~mJy (see Appendix~A). The difference between the LTE and non-LTE spectra, and difference between the spectra computed for different methanol molecular data, are similar to those in the case of the `laminar' disc model. In Table~\ref{tab:lines} we present the list of the 10 strongest CH$_3$OH lines obtained in the non-LTE case for the `fast' mixing model and the `$v_t=0$' scheme. The uncertainty of the absolute values of synthetic line fluxes given in Table~\ref{tab:lines} can be up to 3~mJy. This value was estimated by comparing the methanol spectra obtained with the \textsc{lime} grid consisting of 10\,000 points and the spectra obtained with denser grid of 20\,000 points (see Appendix~A).

\begin{figure*}
\centering
\includegraphics[scale=0.8]{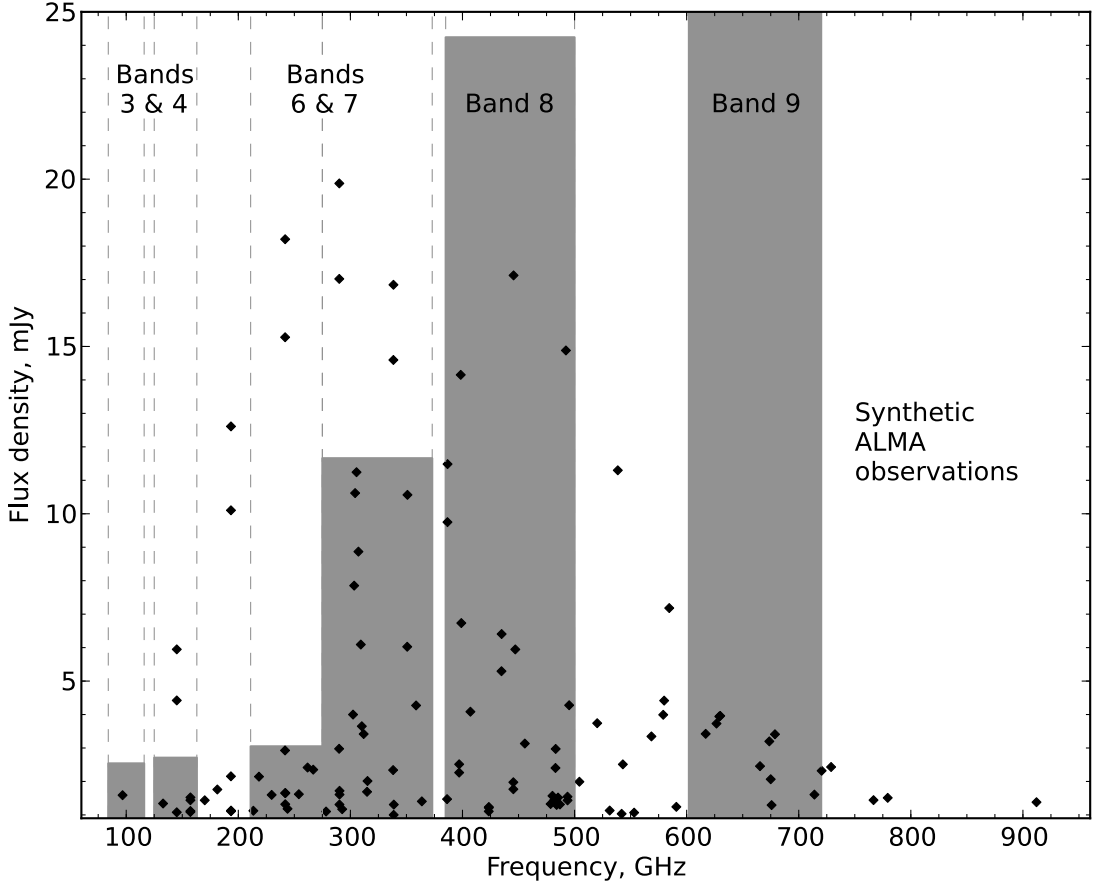}
\caption{The non-LTE peak flux densities of methanol obtained for the disc located at the
distance of 140~pc and a disc inclination angle of $35\degr$. The methanol abundances
were computed with the `fast' mixing disc chemical model. The `$v_t=0$' methanol level
scheme was used. Only lines with peak flux densities $>1$~mJy are shown.
The vertical shaded boxes show the ALMA Bands 3, 4, 6, 7, 8, and 9.
The height of these boxes designates the ALMA sensitivity at the corresponding band
for the 1 hour integration time with the full array  of 50 antennas and a
spectral resolution of 0.2~km~s$^{-1}$.}
\label{fig:fastmix}
\end{figure*}

\begin{table}
\caption{The list of the 10 strongest CH$_3$OH lines as it is predicted by our non-LTE
computations with the `fast' mixing disc chemical model and the `$v_t=0$' methanol level
scheme. The line flux densities were computed assuming that the disc is located at the
distance of 140~pc and an inclination angle of $35\degr$.
The lines that could be good candidates for observational searches with ALMA are
highlighted with the superscript *.}
\label{tab:lines}
\begin{tabular}{@{}cccc}
\hline
Frequency, & Transition & Line peak & Integrated\\
 GHz & & flux density & flux,\\
  & & mJy & mJy~km~s$^{-1}$\\ \hline
$290.111^*$ & $6_0-5_0~A^+$ & 20.9 & 14.6\\
$241.791^*$ & $5_0-4_0~A^+$ & 18.2 & 14.4\\
$445.571$ & $3_1-2_0~A^+$ & 17.1 & 14.0\\
$290.070^*$ & $6_{-1}-5_{-1}~E$ & 17.0 & 12.9\\
$338.409^*$ & $7_{0}-6_{0}~A^+$ & 16.8 & 13.3\\
$241.767^*$ & $5_{-1}-4_{-1}~E$ & 15.3 & 13.7\\
$492.279$ & $4_1-3_0~A^+$ & 14.9 & 12.5\\
$338.345$ & $7_{-1}-6_{-1}~E$ & 14.6 & 12.4\\
$398.447$ & $2_1-1_0~A^+$ & 14.1 & 11.0\\
$193.454$ & $4_0-3_0~A^+$ & 12.6 & 11.3\\
\hline
\end{tabular}
\end{table}

Among the lines listed in Table~\ref{tab:lines}, only the $3_1-2_0~A^+$ (445.571~GHz), $4_1-3_0~A^+$ (492.279~GHz), $2_1-1_0~A^+$ (398.447~GHz) lines show variations of more than $40\%$ due to changes in the A-coefficients. The variations of all other lines from Table~\ref{tab:lines} due to changes in the A-coefficients are small and do not exceed the Monte-Carlo noise level of 0.02~mJy.

The line flux densities obtained with the `fast' mixing model are higher by a factor of 30 than those calculated with the `laminar' disc model. This is mainly due to higher methanol abundances and column density in the `fast' mixing model. Moreover, the ratios of strong lines are different in these two disc chemical models. For example, the ratio of the $7_{0}-6_{0}~A^+$ (338.409~GHz) and $5_0-4_0~A^+$ (241.791~GHz) lines changes from 0.78 in the `laminar' case to 0.93 in the `fast' mixing case. While the \textsc{lime} grids adopted for the `fast' mixing and `laminar' models have the same number of grid points, but different spatial distributions, this cannot lead to such deviations in the line ratios (see Appendix~A).

An increase in the CH$_3$OH abundance also does not lead to the changes in line ratios. We found that the difference in the line ratios is related to the spatial variations of the gas-phase methanol abundance distributions in the `fast' mixing and `laminar' model. As we mentioned in the Section~\ref{sec:transport_model} presenting the disc chemical model, the location of the methanol molecular layer differs between the `laminar' and `fast' mixing chemical models (see Fig.~\ref{fig:methanol_abunds}). Consequently, the line emission presumably forms in different disc regions with different physical conditions.

\begin{figure}
\centering
\includegraphics[scale=0.42]{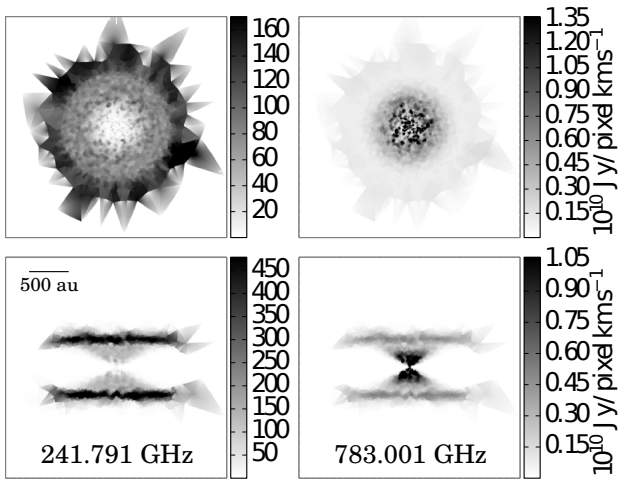}
\caption{The $0$-th moment maps obtained for the disc seen face-on (upper panels) and edge-on (lower panels) for the $5_0-4_0~A^+$ (241.791GHz, left panels) and $3_0-2_0~A^+$ (783.001~GHz, right panels) methanol lines.}
\label{fig:images_inclinations}
\end{figure}

This indicates that the strongest methanol lines and their ratios are sensitive to the physical conditions in the disc. The lines with frequency $>600$~GHz  trace presumably inner disc regions with $r<400$~au, while the lines with frequency $<400$~GHz trace mostly outer disc regions with $r>400$~au (see e.g. Fig.~\ref{fig:images_inclinations}).

To extract the best candidate lines to be detected with ALMA, we compared the line peak flux densities modelled using the `fast' mixing chemical model with the sensitivity after 1~hour of integration time with the full ALMA array (50 antennas), and spectral resolution of 0.2 km~s$^{-1}$. The sensitivity was estimated with the ALMA Sensitivity Calculator for the central frequencies of the ALMA Bands 3, 4, 6, 7, 8 and 9 and shown in Fig.~\ref{fig:fastmix}. We stress that Fig.~\ref{fig:fastmix} is mostly for clarity and should only be used for rough estimation of ALMA on-source integration time. To make reliable estimates of integration time considering nearby discs, we performed simulation of the disc observations with ALMA in Common Astronomy Software Applications (\textsc{casa}) package (see Section~\ref{sec:simobs}).

Among the lines listed in Table~\ref{tab:lines}, the best candidates for searches with ALMA are the $5_0-4_0~A^+$ (241.791~GHz) and $5_{-1}-4_{-1}~E$ (241.767~GHz) methanol lines. Other promising candidates are the $6_0-5_0~A^+$ (290.111~GHz), $6_{-1}-5_{-1}~E$ (290.070~GHz) and $7_{0}-6_{0}~A^+$ (338.409~GHz) methanol lines.

The line fluxes presented in Table~\ref{tab:lines} and the spectrum shown in Fig.~\ref{fig:fastmix} were obtained for the disc inclination angle of $35\degr$. We also studied how the line fluxes change with the disc orientation. The variations of peak line flux densities due to changes in the disc inclination angle are substantial. For example, the peak flux density of $5_0-4_0~A^+$ methanol line at 241.791~GHz is 63 and 10~mJy for the inclination angles of 0 (face-on) and $90\degr$ (edge-on), respectively.

\subsubsection{Simulating ALMA observations}
\label{sec:simobs}

To estimate the on-source integration time needed for detection and imaging of the methanol emission we computed simulated ALMA observations with \textsc{simobserve} and \textsc{simanalyze} tasks of \textsc{casa} package version 4.1.5 (see Appendix~B for the input \textsc{casa} parameters). As input \textsc{casa} models we used the disc images in the $5_0-4_0~A^+$ methanol line (241.791~GHz) computed with  \textsc{lime} for the `fast' mixing disc model with 10\,000 grid points. The simulations were performed for the most compact antenna configuration of the full ALMA array. This configuration was chosen to provide maximum sensitivity. The precipitable water vapor was assumed to be 0.5 mm. The noisy images computed with \textsc{simobserve} were cleaned using natural weighting.

We found that due to low surface brightness the image reconstruction is impossible even after 9 hours of integration for a disc with any inclination located at a distance of 140 pc. The $3\sigma$ unresolved detection of methanol emission is also impossible in a reasonable amount of time ($\sim 2$--$3$~hours of integration).

As the detection of the methanol lines is impossible in the case of DM~Tau disc we considered the simple modifications of our initial model images that can improve the detectability of the CH$_3$OH lines. These modifications may be a good first approximation to study the effects of a distance to a disc, disc size and methanol abundance on the results of simulated observations taking into account relatively large uncertainties in the predicted methanol line intensities.

\begin{figure*}
\centering
\includegraphics[scale=0.41]{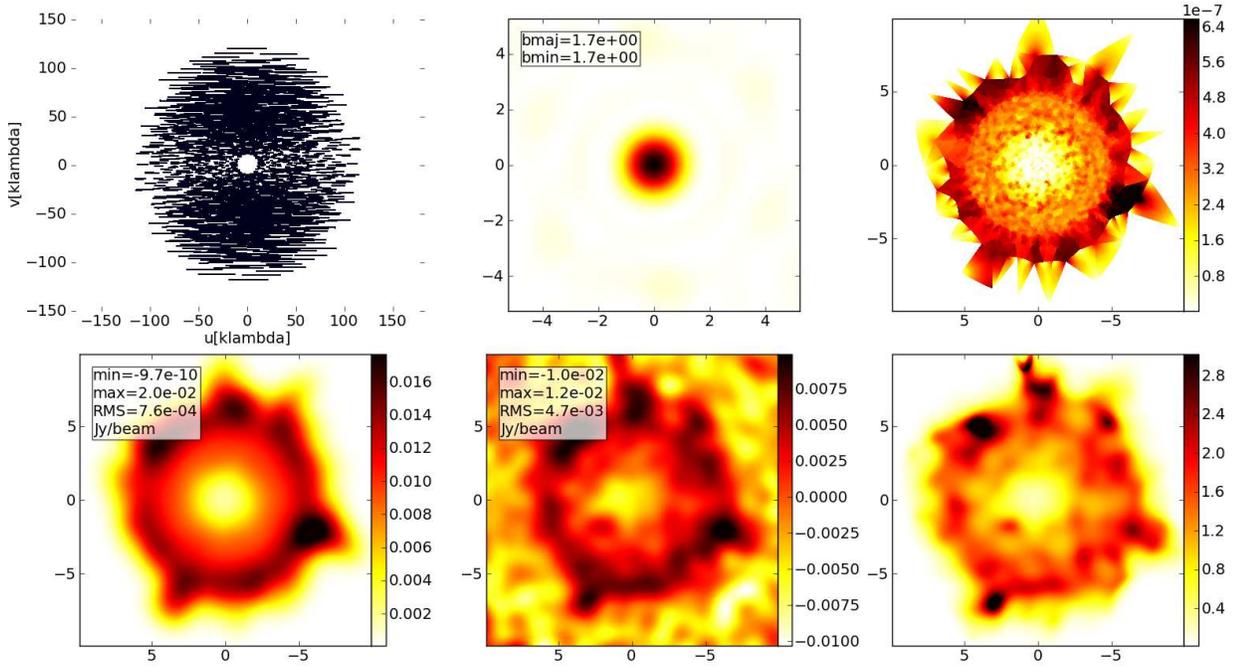}
\caption{The results of disc simulated observations after 3 hours of integration with the full ALMA array. The simulations were performed with \textsc{casa} package for the face-on disc located at a distance of 140 pc. The surface brightness of the input \textsc{casa} model image was increased on a factor of 10 comparing with the one that was obtained using \textsc{lime} for the `fast' mixing disc model. From left to right and from top to bottom: $uv$-coverage, synthesized beam, the input model image, the input image convolved with the beam, the clean synthetic image, fidelity. The image axes, which are not labelled, are in arcsec. The color bars denote intensity in Jy~beam$^{-1}$ in all images except the fidelity image.}
\label{fig:images_analysis}
\end{figure*}

We increased the disc surface brightness by a factor of 10 that is within the possible uncertainties of methanol abundance due to uncertainties in chemical reaction rates. In this case one can see some signs of the ring-like methanol emission for a disc seen face-on (see Fig.~\ref{fig:images_analysis}) after 3 hours of integration. But in general the detection of the methanol emission is very tentative especially for the inclined disc. One can go further and increase the initial disc surface brightness by a factor of 100 that is comparable with the difference of the methanol column density between SW11 and \citet{Walsh_ea14} chemical models. In this case the CH$_3$OH emission can be detected with SNR~$>5\sigma$ and imaged with fidelity\footnote{Fidelity indicates how well the simulated output image $I$ matches the convolved input image $T$ \citep{Pety2001}. Fidelity$=|I|/|I-T|$.}~$\sim10$ after 3 hours of integration for any disc inclination.

\begin{figure*}
\centering
\includegraphics[scale=0.41]{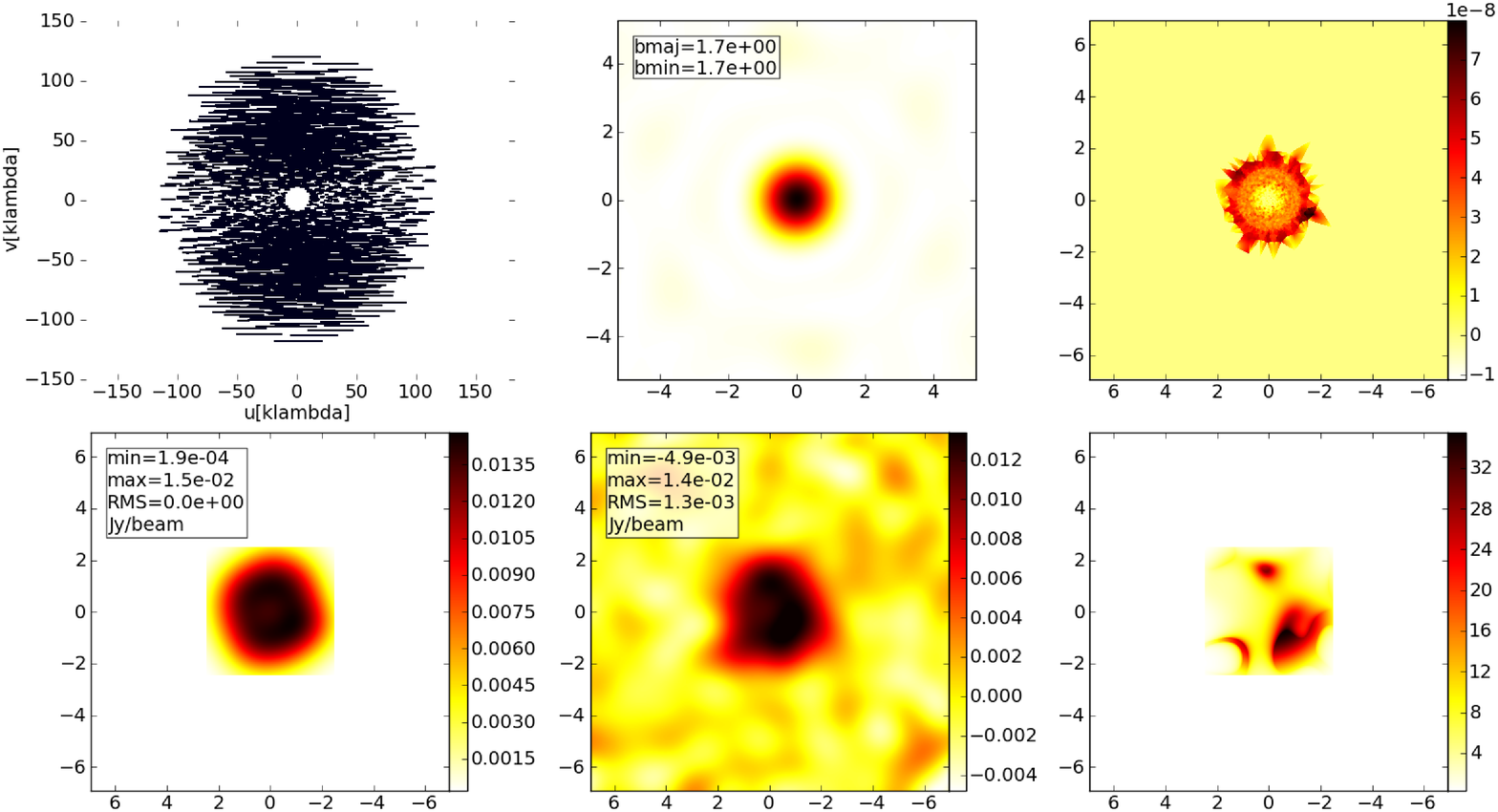}
\caption{The results of disc simulated observations after 3 hours of integration with the full ALMA array. The simulations were performed for the face-on disc located at a distance of 140 pc. The pixel size of the \textsc{casa} input model image was decreased from 0.01 to 0.0025 arcsec comparing with the initial image obtained for the `fast' mixing model. From left to right and from top to bottom: $uv$-coverage, synthesized beam, the input model image, the input image convolved with the beam, the clean synthetic image, fidelity. The image axes, which are not labelled, are in arcsec. The color bars denote intensity in Jy~beam$^{-1}$ in all images except the fidelity image.}
\label{fig:images_analysis1}
\end{figure*}

The angular size of the DM~Tau disc is large comparing with the ALMA beam which makes methanol lines detection more difficult. To simulate the observations of a more compact ($\sim200$ au) disc located at a distance of 140 pc we decreased the pixel size in the initial disc images from 0.01 to 0.0025 arcsec. In this case the large ALMA beam and the disc image (see Fig.~\ref{fig:images_analysis1}) have comparable size. The CH$_3$OH emission can be detected with SNR~$\sim5\sigma$ and imaged (with relatively low spatial resolution) with fidelity~$\sim10$ after 3 hours of integration for the face-on disc. For the edge-on disc, SNR and fidelity after 3 hours of integration are of $\sim3\sigma$ and~$\sim5$, respectively.

\begin{figure*}
\centering
\includegraphics[scale=0.41]{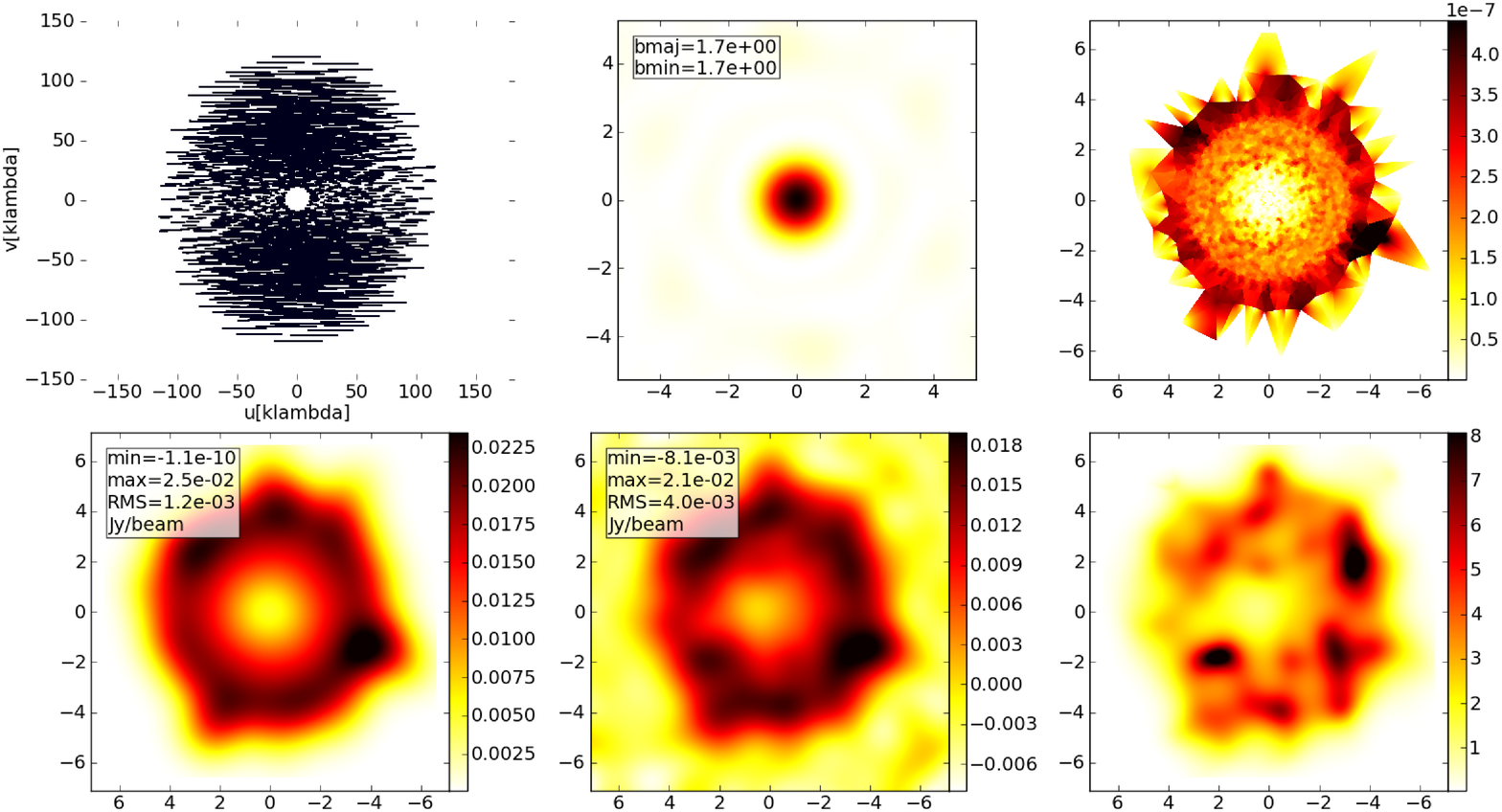}
\caption{The results of  simulated observations obtained for the disc similar to the one of TW~Hya after 3 hours of integration with full ALMA array. From left to right and from top to bottom: $uv$-coverage, synthesized beam, the input model image, the input image convolved with the beam, the clean synthetic image, fidelity. The image axes, which are not labelled, are in arcsec. The color bars denote intensity in Jy~beam$^{-1}$ in all images except the fidelity image.}
\label{fig:images_analysis_TWHya}
\end{figure*}

We also considered the case with the disc similar to the one around Tw~Hya that is closer and smaller than the disc around DM~Tau. The disc around TW~Hya is located at a distance of 54 pc and its radius and inclination is of $\sim200$~au and $6\deg$, respectively \citep{Andrews12}. To simulate observations of this disc we scaled the surface brightness of the \textsc{lime} output images of the face-on disc by a factor of $(140/54)^2$ and decreased the pixel size from 0.01 to 0.0065 arcsec. As a result we obtained  that the methanol emission can be detected with $\sim5\sigma$ SNR and imaged with fidelity~$\sim10$ (see Fig.~\ref{fig:images_analysis_TWHya}) after 3 hours of integration for any disc inclination.

\section{Discussion}

\subsection{Importance of the adopted disc chemical structure}
\label{sec:diss:chem_model}

The largest ambiguity in the analysis of the observability of methanol in protoplanetary discs with ALMA is related to the adopted chemical structure. First, as it was demonstrated by \cite{Vasyunin_ea08}, the lack of accurate chemical gas-phase kinetics data makes calculated abundances of simple COMs in discs, such as e.g. H$_{2}$CO and CH$_{3}$OH, uncertain by a factor of $>3$--$5$. A second source of intrinsic uncertainties is our limited knowledge of grain surface chemical processes, which are believed to be a main synthesis pathway for organic species in space \citep[e.g.,][]{Garrod_Herbst06,2009ARA&A..47..427H}. The SW11 surface chemistry network is adopted from \citet{Garrod_Herbst06,2008ApJ...682..283G} and includes key surface reactions forming and destroying methanol, which are based on accurate laboratory measurements, and best available reaction data. Since no statistical analysis of the impact of the surface chemistry uncertainties on the results of astrochemical modelling has been attempted so far, we roughly estimate that the total intrinsic uncertainties of the calculated methanol abundances in SW11 models adopted in our study are of the order of a factor of $10$--$30$. All other gas-grain astrochemistry disc models should have similar uncertainties for CH$_{3}$OH and other COMs.

Another important factor that controls the disc chemistry is the adopted disc physical structure, including the density and temperature distributions and dust properties. This makes meaningful comparison with other models difficult, even in the case when adopted chemical networks are similar. \citet{Walsh_ea14} have performed a detailed comparison of their results with other contemporary models that have similar chemical complexity, including the `laminar' model adopted in our study \citep[see][]{Willacy_07,SW11}. They have also included the modelling results obtained with two different chemical networks, namely, based on the Ohio State University (OSU, \citet{2008ApJ...682..283G}) and `RATE06' UDFA \citep{Woodall07} databases.

Walsh et~al. have found that our `laminar' model has the lowest CH$_3$OH column density among the other models by a factor of $\sim 10^{3}$, even though the column densities of other key species like CO, H$_{2}$CO and cyanopolyynes are the same. This discrepancy is larger than the intrinsic uncertainties of the SW11 chemical model. The likely cause is a slow grain-surface recombination rate attributed to low diffusivity of surface reactants assumed by SW11, with a ratio of the diffusion to desorption energies of 0.77 \citep{Ruffle_Herbst00}. In contrast, \citet{Walsh_ea14} and \citet{Willacy_07} have considered fast diffusivity of the surface reactants, with a diffusion to desorption energy ratio of 0.3.

Thus, our `laminar' disc model represents a very pessimistic case when one aims at estimating the observability of methanol in quiescent protoplanetary discs with limited grain surface chemistry. Not surprisingly, 3D non-LTE line radiative transfer simulations with such a disc model predict that no methanol lines can be detected with the full ALMA within a reasonable amount of time ($\sim 2$--$3$~hours of integration per source).

By contrast, the column density of methanol in our dynamically active (`fast' mixing) disc model is higher by about two orders of magnitude compared to the `laminar' model. As was mentioned in Section~\ref{sec:transport_model}, this is an effect of transport of ices from disc midplane to elevated heights where they can be more easily photodesorbed and photoprocessed, and from cold disc regions to warm disc regions, which facilitates the surface synthesis of complex species (see discussion about organics in SW11).

While the CH$_3$OH column density in the `fast' mixing model is still lower by an order of magnitude compared to the values from other chemical models, the difference is within the intrinsic chemical uncertainties for methanol. Thus, our `fast' mixing model represents a more optimistic case for methanol line detection in discs. Consequently, with the `fast' mixing model, our calculations predict that several methanol lines can be detected with the ALMA (see Table~\ref{tab:lines}). The two best candidates from our study are the CH$_{3}$OH~$5_0-4_0~A^+$ (241.791~GHz) and $5_{-1}-4_{-1}~E$ (241.767~GHz) lines, which can possibly be detected with $\sim 5 \sigma$ SNR within $\sim 3$~hours of integration with the full ALMA in the direction of the disc similar to the one of TW Hya.

Please note however that these lines are different from those that were identified by LTE LRT analysis by \citet{Walsh_ea14}. Their best detection candidates are $3_1-3_0~A^{-+}$ (305.474~GHz), $2_1-1_0~A^+$ (398.447~GHz) and $5_{-2}-4_{-1}~E$ (665.442~GHz) lines. According to our non-LTE calculations, these methanol lines can not be detected in nearby face-on discs even with the full ALMA within two hours of integration. This can explain why Walsh et~al. were not able to detect these lines in several bright discs with their ALMA Cycle~0 and 1 campaigns.

The difference between our and \citet{Walsh_ea14} candidate lines is not only due to difference between predictions of chemical models but also due to non-LTE effects. According to our calculations non-LTE flux densities for the line at 665.442~GHz are lower by a factor of $\sim25$ than LTE flux densities that is comparable with the difference in methanol abundance predicted by SW11 'fast' mixing and \citet{Walsh_ea14} chemical models. The LTE peak flux density computed for this line with 'fast'
 mixing model exceeds the estimated ALMA sensitivity ($\sim$30 mJy for 1 hour integration time) by a factor of 2 which makes this line a good candidate, opposing the results of non-LTE LRT calculations.

\citet{Walsh_ea14} have also estimated the line fluxes for those transitions that are identified as promising for ALMA searches by us. They estimated the integrated flux density of $5_0-4_0~A^{+}$ (241.79~GHz) and $7_{0}-6_{0}~A^{+}$ (338.409~GHz) lines to be 71 and 31 mJy~km~s$^{-1}$, respectively. These values are significantly higher than those found in our study.

\section{Conclusions}

We performed line radiative transfer computations for a T~Tauri protoplanetary disc model assuming non-LTE excitation of energy levels, and estimated methanol line fluxes in the (sub-)millimetre range available to ALMA. The disc chemical model used in this study has rather low methanol column densities among other modern disc chemical models. Thus, the line fluxes computed in this study are on the pessimistic side. With our calculations we found the following key results:\\
- The difference between LTE and non-LTE methanol line flux densities can be as large as a factor of $10$--$50$;\\
- High-frequency lines ($>600$~GHz) trace presumably inner disc regions while low-frequency lines ($<400$~GHz) trace outer disc regions;\\
- The ratio of the strongest methanol lines is sensitive to the physical conditions in the disc;\\
- The best candidates for observational searches with ALMA are the $5_0-4_0~A^+$~(241.791~GHz) and $5_{-1}-4_{-1}~E$ (241.767~GHz) methanol lines. These lines can be detected with 5$\sigma$ SNR within about 3 hours of integration with the full ALMA array (50 antennas) for nearby compact discs located at a distance $\la 140$~pc;\\
- The other good candidates for observations with ALMA are $6_0-5_0~A^+$ (290.111~GHz), $6_{-1}-5_{-1}~E$ (290.070~GHz) and $7_{0}-6_{0}~A^+$ (338.409~GHz) lines;\\
- Schemes of the methanol levels and transitions similar to those from the LAMDA database (without levels of the torsionally excited states and collisions with helium) allow us to obtain reliable flux density estimates for methanol lines in protoplanetary discs around low-mass stars; \\
- The inclusion of collisions with helium atoms can affect the predicted methanol line ratios, especially for pairs of lines with frequencies separated by more than $\sim 200$~GHz;\\
- There are no bright methanol masers predicted for the T~Tauri protoplanetary disc model.

\section*{Acknowledgements}
S. Parfenov and A. Sobolev work in Ural Federal University with financial support from the Russian Science Foundation (project no. 15-12-10017). D. Semenov acknowledges support by the {\it Deutsche Forschungsgemeinschaft} through SPP~1385: ``The first ten million years of the solar system - a planetary materials approach'' (SE 1962/1-3). This research made use of NASA's Astrophysics Data System. This research made use of \textsc{aplpy}, an open-source plotting package for \textsc{python} hosted at http://aplpy.github.com. Most of figures in this paper were constructed with the \textsc{matplotlib} package \citep{Hunter07}.

This work used the DiRAC Data Centric system at Durham University, operated by the Institute for Computational Cosmology on behalf of the STFC DiRAC HPC Facility (www.dirac.ac.uk). This equipment was funded by a BIS National E-infrastructure capital grant ST/K00042X/1, DiRAC Operations grant ST/K003267/1 and Durham University. DiRAC is part of the National E-Infrastructure. 

\bibliographystyle{mn2e}

\section*{Appendix A}
One of the main features of the Monte-Carlo calculations is the presence of residual noise. The minimum signal-to-noise ratio of the level populations for our calculations estimated according to \citet{2010A&A...523A..25B} is about $1.5\times10^8$. One of many possible ways to estimate the noise level in synthetic spectra is to compute a relatively large number of spectra with the same model parameters and to estimate their standard deviation. Since such computations require too prohibitive an amount of CPU time, we had to use another approach. 

\begin{figure}
\centering
\includegraphics[scale=0.46]{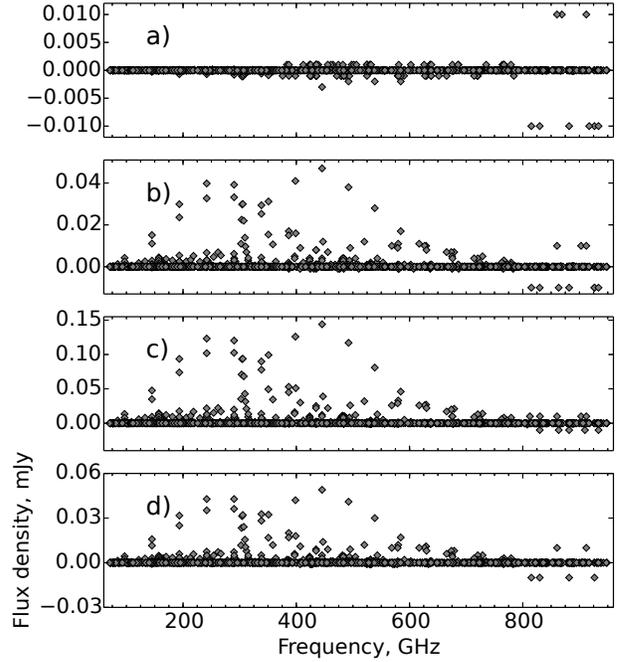}
\caption{The difference between the spectra obtained with the `laminar' and
`$v_t=0$' methanol level scheme. \textbf{a)} The difference obtained after two successive
\textsc{lime} runs with the same model parameters. \textbf{b)} The difference obtained after
two successive \textsc{lime} runs with the same model parameters, but two realisations
of the 10\,000-point \textsc{lime} grid. \textbf{c)} The difference obtained with two different
\textsc{lime} grids. The first grid includes 10\,000 points within the disc and 10\,000 sink points. The second grid includes 20\,000 point within the disc and 20\,000 sink points.
\textbf{d)} The difference obtained with two different
\textsc{lime} grids. The first grid includes 10\,000 points within the disc and 10\,000 sink points. The second grid includes 40\,000 point within the disc and 40\,000 sink points.
}
\label{fig:noise_multi}
\end{figure}

To estimate the Monte-Carlo noise level in our LRT simulations, we compared the spectra computed with three successive \textsc{lime} runs (using the same model parameters). The spectra were computed with the same \textsc{lime} grid and the `$v_t=0$' methanol level scheme. As an example, the difference between the two spectra computed for the `laminar' model is presented in Fig.~\ref{fig:noise_multi}, panel a. This difference is random and it is maximum for transitions with frequencies higher than 800~GHz. It means that such transitions are sensitive to uncertainty in level populations.

The difference between the methanol spectra computed with three successive \textsc{lime} LRT runs for the `laminar' model does not exceed 0.01~mJy. Hence, the value of 0.01~mJy was considered as the Monte-Carlo noise level of the spectra computed with the `laminar' disc model. Using the same approach, we estimated the Monte-Carlo noise level for the `fast' mixing disc model, with the value of 0.02~mJy.

The uncertainty in the synthetic spectra can also be due to randomness of the spatial distribution of the grid points in the randomly generated \textsc{lime} grid. To estimate this uncertainty, we computed the three synthetic methanol spectra with the same model parameters but with three realisations of the \textsc{lime} grid. As an example, the difference between the two spectra computed for the `laminar' disc chemical model is shown in Fig.~\ref{fig:noise_multi}, panel b. The difference is systematic and for most transitions is proportional to the line intensity.

The difference between the three spectra computed for the `laminar' model does not exceed the value of 0.05~mJy. We consider the flux density value of 0.05~mJy as an estimate of the uncertainty of the modelled flux densities due to randomness of the \textsc{lime} grid in the case of the `laminar' disc model. For the `fast' mixing disc chemical model this uncertainty was calculated to be 1~mJy.

The results can also depend on the total number of the \textsc{lime} grid points. The results presented in Section~\ref{sec:results} were obtained with the basic \textsc{lime} grids that includes 10\,000 points within the disc model and 10\,000 sink points. To study how the LRT results can be affected by the choice of these numbers, we performed \textsc{lime} calculations with a denser grid that includes 20\,000 points within the disc model and 20\,000 sink points.

In Fig.~\ref{fig:noise_multi}c we present the difference between the methanol line peak flux densities computed with the 10\,000-point grid and with the denser, 20\,000-point grid, using the `laminar' disc model. As can be clearly seen from Fig.~\ref{fig:noise_multi}, panel c, the difference between the spectra computed with the two different grids does not exceed 0.2~mJy. In the case of the `fast' mixing disc model this difference does not exceed 3.0~mJy. In Fig.~\ref{fig:noise_multi}d we demonstrate the difference between spectra computed with 10\,000 and 40\,000-point grids (see Section~\ref{sec:lrt_calc} for the details on the 40\,000-point grid). The difference between these spectra again does not exceed 0.2 and 3.0 mJy for the `laminar' and `fast' mixing disc models, respectively.

\begin{figure}
\centering
\includegraphics[scale=0.46]{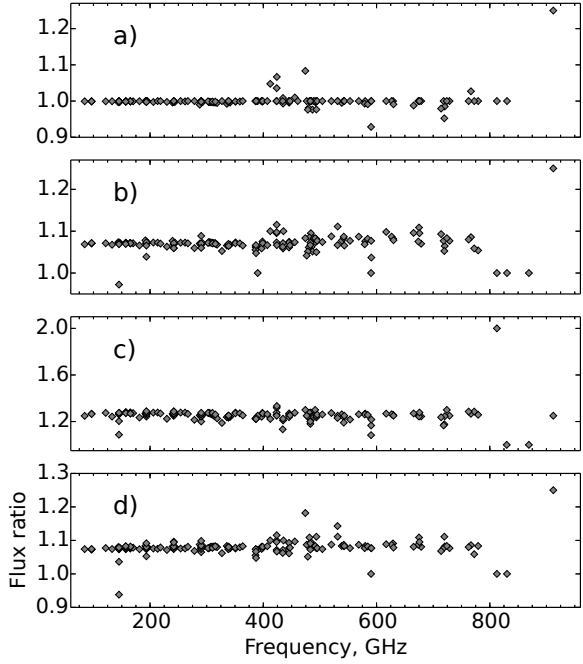}
\caption{The ratios between the spectra obtained with the `laminar' disc model and
`$v_t=0$' methanol level scheme. The ratios are shown only for the lines with peak flux
densities that exceed 0.01~mJy. \textbf{a)} The ratio obtained after two successive
\textsc{lime} runs with the same model parameters. \textbf{b)} The ratio obtained after
two successive \textsc{lime} runs with the same model parameters, but two realisations
of the 10\,000-point \textsc{lime} grid. \textbf{c)} The ratio obtained with two different
\textsc{lime} grids. The first grid includes 10\,000 points within the disc and 10\,000 sink points. The second grid includes 20\,000 point within the disc and 20\,000 sink points.
\textbf{d)} The ratio obtained with two different
\textsc{lime} grids. The first grid includes 10\,000 points within the disc and 10\,000 sink points. The second grid includes 40\,000 point within the disc and 40\,000 sink points.
}
\label{fig:noise_multi_ratio}
\end{figure}

It is important to point out that the Monte-Carlo noise and variations in the grid parameters do not lead to significant systematic variations in the relative line flux densities, i.e. high-frequency lines do not become significantly more intense or less intense compared to the low-frequency lines (see Fig.~\ref{fig:noise_multi_ratio}). Most of the lines that demonstrate large relative variations due to the Monte-Carlo noise and variations of the grid parameters are relatively weak and are negligible from the point of view of observations.

\section*{Appendix B}

The following is an example of the script used to simulate ALMA observations of the disc with \textsc{casa} 4.5.1 package.

\noindent\textsl{default("simobserve") \\
skymodel = "image.fits" \\
integration = '10s' \\
incell	       = "0.01arcsec" \\
incenter       = "241.791GHz" \\
inwidth        = "0.1612MHz" \\
setpointings   = False \\
ptgfile        = "pointing.txt" \\
maptype        =     'ALMA' \\
antennalist    = "alma.out01.cfg" \\
obsmode	       = "int" \\
refdate        = "2016/11/23" \\
hourangle      =  "transit" \\
totaltime      =  "3h" \\
thermalnoise   =      "tsys-atm" \\
t\_ground       =      269.0 \\
simobserve() \\
\#\\
default("simanalyze") \\
niter              = 200000 \\
threshold          = "0.0041Jy/beam" \\
weighting          = "natural" \\
analyze = True \\
showconvolved = True \\
simanalyze() \\
}

\noindent The "pointing.txt" file contains the following line: \\
\noindent J2000 00:00:00.000 +000.00.00.000000  10.0

\label{lastpage}

\end{document}